\documentclass{aa}
\usepackage{epsfig}

\def\simless{\mathbin{\lower 3pt\hbox
     {$\rlap{\raise 5pt\hbox{$\char'074$}}\mathchar"7218$}}} 
\def\simgreat{\mathbin{\lower 3pt\hbox
     {$\rlap{\raise 5pt\hbox{$\char'076$}}\mathchar"7218$}}} 

\newcommand{\Lsun} {L$_\odot$}

\begin{document}

\title{Mid--infrared emission of galactic nuclei\thanks{Based on ESO:
68.B-0066(A) and observations with ISO, an ESA project with
instruments funded by ESA Member States (especially the PI countries:
France, Germany, the Netherlands and the United Kingdom) with the
participation of ISAS and NASA.}:}

\subtitle{TIMMI2 versus ISO observations and models}

\author {R.~Siebenmorgen\inst{1} \and E.~Kr\"ugel\inst{2}
         \and H.W.W. Spoon\inst{3}}

\institute{
        European Southern Observatory, Karl-Schwarzschildstr. 2, 
        D-85748 Garching b. M\"unchen, Germany 
\and
        Max-Planck-Institut for Radioastronomy, Auf dem H\"ugel 69,
	Postfach 2024, D-53010 Bonn, Germany 
\and
	Kapteyn Institute, P.O. Box 800, NL-9700 AV Groningen, 
        The Netherlands
}

\offprints{rsiebenm@eso.org}
\date{Received XXXX   / Accepted    XXXX}

\abstract {We investigate the mid--infrared radiation of galaxies that
are powered by a starburst or by an AGN.  For this end, we compare the
spectra obtained at different spatial scales in a sample of infrared
bright galaxies.  ISO observations which include emission of the
nucleus as well as most of the host galaxy are compared with TIMMI2
spectra of the nuclear region.  We find that ISO spectra are generally
dominated by strong PAH bands.  However, this is no longer true when
inspecting the mid--infrared emission of the pure nucleus.  Here PAH
emission is detected in starbursts whereas it is significantly reduced
or completely absent in AGNs.  A physical explanation of these new
observational results is presented by examining the temperature
fluctuation of a PAH after interaction with a photon.  It turns out
that the hardness of the radiation field is a key parameter for
quantifying the photo--destruction of small grains.  Our theoretical
study predicts PAH evaporation in soft X--ray environments.  Radiative
transfer calculations of clumpy starbursts and AGN corroborate the
observational fact that PAH emission is connected to starburst
activity whereas PAHs are destroyed near an AGN.  The radiative
transfer models predict for starbursts a much larger mid--infrared
size than for AGN. This is confirmed by our TIMMI2 acquisition images:
We find that the mid--infrared emission of Seyferts is dominated by a
compact core while most of the starbursts are spatially resolved.
\keywords{      Infrared: galaxies -- 
                Galaxies: ISM --
                Galaxies: nuclei --
                Galaxies: dust         }  }

\maketitle

\section{Introduction}
Luminous infrared galaxies in the local universe with infrared
luminosities exceeding 10$^{11}$ \Lsun\ were discovered in the IRAS
all sky survey (Soifer et al.~1987).  Responsible for their huge
luminosity may be a combination of starburst and AGN activity.  In
both cases, the nuclear emission is fuelled by an enormous
concentration of gas and dust in the central region (Sanders \&
Mirabel 1996).

Luminous infrared galaxies are also found to be a major component in
the early universe ($z \simgreat 1$).  Deep mid--infrared surveys with
the infrared space observatory, ISO, could identify individual objects
which account for most of the cosmic infrared background light.  It is
argued that the emission of most of the distant infrared galaxies
comes mainly from starbursts (Elbaz et al.~2002).  This hypothesis may
also be tested by inspecting the spectra of local galaxies.  ISO
spectra of local galaxies (Rigopoulou et al.~1999, Siebenmorgen et
al.~1999, Clavel et al.~2000, Laureijs et al.~2000, Laurent et
al. 2000) usually show a set of infrared emission bands which are
attributed to the emission of polycyclic aromatic hydrocarbons (PAH).

The question as to which source, on average, is responsible for the
luminosity of the cosmological background is still a matter of intense
debate and, at the moment, different answers are given for different
wavelength regions.  The X--ray background could be partly resolved by
ROSAT into individual components and Hasinger et al.~(1998) showed
that at X--rays AGNs make the greatest contribution.  At infrared
wavelengths, on the other hand, starbursts seem to dominate, as
infered from numerous ISO observations (Genzel \& Cesarsky 2000).

We want to re-address the question whether infrared bright galaxies are
predominantly powered by dust-enshrouded AGN or starbursts.  As a new
observational technique, we employ mid--infrared sub--arcsec imaging
and mid--infrared spectroscopy at various spatial scales.  (For
previous ground based mid--infrared spectroscopy see for example:
Smith et al.~1989, Roche et al.~1991 and Soifer et al.~2002). The
mid--infrared wavelength region is of particular interest as the
optical depth is usually sufficiently low to penetrate the dust
enshrouded nucleus.  Modern mid--infrared cameras have adequate
spatial resolution to estimate the size of the emitting regions which
is an important parameter to constrain the nature of the underlying
source: At comparable distances of the sources a resolved
mid--infrared emission points towards a starburst galaxy while an
unresolved nucleus indicates an AGN or an extreme compact star
cluster.  As we will demonstrate in model calculations, infrared
emission bands are a diagnostic tool to derive the activity type.  The
ratio of AGN versus starburst luminosity of local galaxies may be
quantified by mid infrared spectra observed at different spatial
resolutions.

The paper is structured as follows: In Sect.~2 we introduce the
sample of infrared bright galaxies observed with ISO and TIMMI2.  The
observation are summarised in Sect.~3, the data analysis procedure
is explained in Sect.~4.  Observational results of individual
galaxies are given in Sect.~5.  Sect.~6 and 7 presents radiative
transfer calculations, which also illustrate the effect of the optical
depth on the SED, and a discussion of the heating and evaporation of
PAH in starburst and AGN environments.

\section {The sample}

The ISO archive contains more than 250 galaxies that have been
spectroscopically observed in the mid--infrared. We have selected 22
of these galaxies for observation with TIMMI2 to compare the large
aperture ISO spectra to the narrow slit TIMMI2 spectra.  The selected
galaxies all have 10\,$\mu$m ISO flux densities above $\sim$100mJy and
are visible from La Silla ($-29^{\rm o}$ South).  Basic properties of
the sample are presented in Table \ref {basic.tab}.  The flux
criterion was prompted by the sensitivity of TIMMI2 in spectroscopic
mode which is $\sim$100mJy/10$\sigma$ in 1 hour.  The sample is
heterogenous with respect to the galaxy type: it contains six
starbursts (SB), one HII, two Sy\,1 and thirteen Sy\,2.  The distances
range from 3.5--176\,Mpc and the luminosities from $0.7 - 120 \times
10^{10}$\Lsun.  Our TIMMI2 sample also contains sources lacking an ISO
spectrum.  These sources were selected from the Markarian catalogue
(Markarian et al. 1989) to fill gaps in our primary observing
schedule.

\begin{table}
\label{basic.tab}
\caption{A few parameters of our IRG sample. Velocities (col.~3) and distances
(col.~4) are from NED.  Col.~5 gives the linear size of 1$''$ at the
distance $D$, and col.~6 the infrared luminosity.  }
\begin{center}
\begin{tabular}{ | l |l |r | l | r | l | }
\hline
                 &      &       &      &   &\\
Name             & Type &    v  & D    & 1$''$  & L   \\
                 &      &       &      &   & \\
                 &      &  km/s &Mpc & pc  & $10^{10}$\Lsun \\
                 &      &       &      &   & \\
\hline
                 &       &      &      &     & \\
Centaurus A      &Sy\,2.0 &547   &   3.5&  17 &  0.68 \\
Circinus         &Sy\,2.0 &439   &   4  &  19 &  1.4  \\
IC4329a          &Sy\,1.2 &4813  &  65  & 315 &  7.6 \\
IRAS05189$-$2524   &Sy\,2.0 &12760 & 176  & 852 &120 \\
IRAS08007$-$6600   &HII     &12324 & 170  & 822 & 34\\
M83              &SB     &516   &   4  &  19 &  0.65 \\
Mrk509           &Sy\,1.2 &10312 & 141  & 684 & 14  \\
Mrk1093          &SB     &4441  &  60  & 290 &  9.3\\
Mrk1466          &SB     &1327  &  18  &  87 &  0.7 \\
NGC1068          &Sy\,2.0 &1148  &  14  &  68 & 19 \\
NGC1365          &Sy\,1.8 &1636  &  22  & 106 &  8.3 \\
NGC1386          &Sy\,2.0 &868   &  18  &  87 &  0.71 \\
NGC2966          &SB     &2044  &  27  & 131 &  1.1 \\
NGC3256          &SB     &2738  &  37  & 178 & 35  \\
NGC4388          &Sy\,2.0 &2524  &  34  & 164 &  1.2 \\
NGC5506          &Sy\,1.9 &1853  &  25  & 120 &  2.3 \\
NGC5643          &Sy\,2.0 &1199  &  16  &  78 &  1.8 \\
NGC6000          &SB     &2193  &  29  & 141 &  9.0 \\
NGC6240          &Sy\,2.0 &7339  & 100  & 483 & 60 \\
NGC7552          &SB     &1585  &  21  & 102 &  9.8 \\
NGC7582          &Sy\,2.0 &1575  &  21  & 102 &  6.3 \\
NGC7674          &Sy\,2.0 &8671  & 118  & 573 & 31 \\
PKS2048$-$57       &Sy\,2.0 &3402  &  46  & 222 &  6.1 \\
                 &       &      &      &     & \\
\hline
\end{tabular}
\end{center}
\end{table}

\section{Observations}

\begin{table}
\caption{Log of TIMMI2 Observations}
\begin{center}
\begin{tabular}{ | l c r l l | }
\hline
            &         &           &          & \\
Target        & Date        & Time       & Filter    & Calibrator \\
              & (2002)       &  (sec)    & ($\mu$m)  &  \\
\hline
            &         &           &          & \\
 Centaurus A      &  Jan. 30&     645   & 10.4     & HD81420\\
            &  Jan. 30&    3225   & grating  & HD81420\\
  Circinus      &  Jan. 30&     268   & 10.4     & HD81420\\
            &  Jan. 30&     967   & grating  & HD81420\\
  IC4329A   &  Jan. 29&     403   & 10.4     & HD81420\\
            &  Jan. 29&    2419   & grating  & HD81420\\
  IRAS05189$-$2524    &  Feb. 01&     361   & 8.6      & HD32887\\
            &  Feb. 01&    2257   & grating  & HD32887\\
            &  25.01.03 & 600  & 10.4     & HD123139 \\
            &  25.01.03 & 600  & 11.9     & HD123139 \\
  IRAS08007$-$6600    &  Feb. 01&     361   & 8.6      & HD32887\\
            &  Feb. 01 &    469    & 11.9  & HD32887\\
            &  Feb. 01&    1935   & grating  & HD32887\\
            &  25.01.03 & 360  & 11.9     & HD123139 \\
  M83       &  Jan. 30&     322   & 10.4     & HD81420 \\
            &  Jan. 30&     301   & 11.9     & HD81420 \\
            &  Jan. 30&    2419   & grating  & HD81420 \\
            &  Jan. 30&     322   & 10.4     & HD81420 \\
  Mrk509    &  Aug. 14&     619   & 11.9     & HD196171\\
            &  Aug. 14&     619   & 11.9     & HD196171\\
            &  Aug. 14&    2580   & grating  & HD196171\\
  Mrk1093   &  Feb. 02&     602   & 11.9     & HD32887\\
            &  Feb. 02&    2903   & grating  & HD32887\\
  Mrk1466   &  Feb. 04&     903   & 8.6      & HD123139\\
            &  Feb. 02&     903   & 11.9     & HD123139\\
            &  Feb. 02&    2419   & grating  & HD123139\\
  NGC1068   &  Jan. 29&     268   & 10.4     & HD32887 \\
            &  Jan. 29&    1209   & grating  & HD32887 \\
            &  Jan. 31&     150   & 8.6      & HD32887 \\
            &  Jan. 31&     604   & grating  & HD32887\\
  NGC1365   &  Jan. 30&     645   & 10.4     & HD32887\\
            &  Jan. 30&    3225   & grating  & HD32887\\
            &  Jan. 31&     150   & 8.6      & HD32887\\
            &  Jan. 31&    1935   & grating  & HD32887\\
  NGC1386   &  Aug. 15&     619   & 11.9     & HD1522\\
            &  Aug. 15&    2580   & grating  & HD1522\\
  NGC2966   &  Feb. 03&     602   & 11.9     & HD123139\\
  NGC3256   &  Jan. 31&     180   & 8.6      & HD123139\\
            &  Jan. 29&     537   & 10.4     & HD81420 \\
            &  Jan. 29&    3628   & grating  & HD81420 \\
            &  Jan. 31&    3064   & grating  & HD123139\\
  NGC4388   &  Feb. 03&     602   & 11.9     & HD123139\\
            &  Feb. 03&    2419   & grating  & HD123139\\
  NGC5506   &  Feb. 02&     602   & 11.9     & HD123139\\
            &  Feb. 02&    1612   & grating  & HD123139\\
  NGC5643   &  Feb. 02&     602   & 11.9     & HD123139\\
            &  Feb. 02&    1612   & grating  & HD123139\\
            &  Aug. 14&     602   & 11.9     & HD123139\\
            &  Aug. 14&    1612   & grating  & HD123139\\
  NGC6000   &  Jan. 29&     215   & 10.4     & HD123139 \\
            &  Feb. 01&     301   & 8.6      & HD123139 \\
            &  Feb. 01&    2096   & grating  & HD123139 \\
  NGC6240   &  Aug. 13&     619   & 11.9     & HD123139\\
            &  Aug. 15&     619   & 11.9     & HD133774\\
            &  Aug. 13&    3096   & grating  & HD123139\\
            &  Aug. 15&    2580   & grating  & HD133774\\
\hline
\end{tabular}
\end{center}
\label{log.tab}
\end{table}

\setcounter{table}{1}
\begin{table}
\caption{continued}
\begin{center}
\begin{tabular}{ | l c r l l | }
\hline
               &            &        &  & \\
Target        & Date        & Time & Filter & Calibrator \\
              & (2002)       &  (sec)    &($\mu$m)         &  \\
\hline
            &         &           &          & \\
  NGC7552   &  Aug. 15&     929   & 11.9     & HD1522\\
            &  Aug. 15&    2580   & grating  & HD169916\\
            &  Aug. 15&    3096   & grating  & HD1522\\
  NGC7582   &  Aug. 14&     619   & 11.9     & HD178345\\
            &  Aug. 14&    2580   & grating  & HD178345\\
            &  Aug. 14&    2709   & grating  & HD1522\\
  NGC7674   &  Aug. 15&     619   & 11.9     & HD178345\\
 PKS2048$-$57 &  Aug. 14&     619   & 11.9     & HD178345\\
            &  Aug. 14&    2460   & grating  & HD178345\\
            &         &           &          & \\
\hline
\end{tabular}
\end{center}
\end{table}

The N band imaging and spectroscopic observations were made with the
TIMMI2 mid--infrared camera (Reimann et al. 2000, K\"aufl et al. 2003)
at the ESO 3.6m telescope.  The camera uses a 240 $\times$ 320 pixel
Raytheon Si:As array. In imaging we selected the 0.2$''$ pixel scale
providing a total field of view of $48'' \times 64''$. For long--slit
spectroscopy the resolving power is $\Delta \lambda / \lambda \sim
400$ and we used the 1.2$''$ or 3$''$ slit and 0.45$''$ pixel field of
view. Table~\ref{log.tab} reports for each galaxy: observing date,
total on--source integration time, filter and calibration standard.
The filter is designated by its central wavelength and the entry {\it
grating} refers to 8--13 $\mu$m spectroscopy.

Observations were performed in chopping and nodding mode with 12$''$
amplitude each with is a compromise between sensitivity and the
ability to measure extended structures.  To avoid saturation of the
detector by the high ambient photon background the elementary
integration times were typically 20\,ms for imaging and 40\,ms for
spectroscopy.  Elementary images were coadded in real time during the
chopping cycles.  Chopping was in North-South direction at a frequency
of $\sim 3$Hz.  After an interval of typically less than a few minutes
the telescope was nodded in order to cancel the telescope emission and
residuals in the sky subtraction. After coaddition of all chopping and
nodding pairs there are two negative and one positive beam (slit) on
the image which we combine applying a shift--and--add procedure. We
note that TIMMI2 is not capable of measuring low surface brightness
emission that is extended by more than half the chopping or nodding
amplitude. Consequently, part of the diffuse emission is cancelled by
the observing technique.

TIMMI2 follow--up observations of ISO galaxies were obtained during
January 28--February 3, August 13--14, 2002 and 4 photometric imaging
observations on January 25, 2003. The observations of the targets were
interleaved with those of nearby mid--infrared calibration standard
stars, typically 4--5 of which were measured each night.  Model
spectra for the standards have an absolute photometric uncertainty of
a few percent (Cohen et al. 1999). However, for stars of spectral type
later than F there might be fundamental molecular lines present in the
photospheric spectra, such as Si--O in the 9$\mu$m region (Heras et
al.  1998). These bands are not considered by the present photospheric
models.  Consequently, one must be careful not to over-interpret faint
structures at these wavelengths. Furthermore close to the blue and red
cut--off frequencies of the grating the terrestrial background varies
strongly with ambient conditions and atmospheric corrections are more
difficult. The same holds for wavelength regions of strong terrestrial
lines near 9.58$\mu$m, 11.73$\mu$m and 12.55$\mu$m.

The N band imaging of the stars was performed in the same filter as
the target and used for photometric flux conversion from photon count
rates into astronomical units (mJy) and to establish the point spread
function (PSF). The spectra of the stars were used for atmospheric
corrections and for flux calibration of the target spectrum.

\section{Data analysis}

\subsection{Mid--infrared imaging}

\begin{table}
\label{flux.tab}
\caption{Mid--infrared flux densities of TIMMI2
observations. Photometric uncertainty $\sim$10\%.}
\begin{center}
\begin{tabular}{ | l | c | c | c  | }
\hline
             &            &        & \\
Name         &F(8.6)      &F(10.4) &F(11.9) \\
             &(Jy)        &(Jy)    &(Jy) \\         
             &            &        & \\
\hline
             &            &        & \\
Centaurus A         &-           &0.65    &-\\
Circinus     &-           &4.10    &-\\
IC4329A      &-           &0.64    &-\\
IRAS05189$-$2524    &0.42        &0.42    &0.57 \\
IRAS08007$-$6600    &0.21        &-       &0.23 \\
Mrk509       &-           &-       &0.22\\
Mrk1093      &-           &-       &0.16\\
Mrk1466      &0.09        &-       &0.11\\
M83          &-           &0.11    &0.23\\
NGC1068      &14.8        &17.6    &-\\
NGC1365      &0.40        &0.46    &-\\
NGC1386      &-           &-       &0.40\\
NGC2966      &-           &-    &$<$0.03\\
NGC3256      &0.45        &0.39    &-\\
NGC4388      &-           &-       &0.29\\
NGC5506      &-           &-       &1.06\\
NGC5643      &-           &-       &0.31\\
NGC6000      &0.21        &0.20    &-\\
NGC6240      &-           &-       &0.28\\
NGC7552      &-           &-       &1.87\\
NGC7582      &-           &-       &0.69\\
NGC7674      &-           &-       &0.26\\
PKS2048$-$57   &-           &-       &0.92\\
             &            &        & \\
\hline
\end{tabular}
\end{center}
\end{table}

We apply multi--aperture photometry on the shift--and--add processed
image, which we refer to as raw or original image.  For each aperture
centred on the brightest pixel of the source we estimate the
background as the mean flux derived in a 6 pixel wide annulus which is
put 6 pixels away from the outer source aperture.  The thus estimated
aperture fluxes, $F(r)$, were plotted against the aperture radii, $r$,
to provide the growth curve. The flux where the growth curve of the
aperture photometry flattens was identified interactively and is taken
as the total flux density.  The absolute photometric uncertainty is
$\sim 10\%$. This error estimate is based on internal consistency of
the calibration observations and monitoring of photometric calibration
standard stars. We note that part of this error is due to large scale
($\geq 10''$) sensitivity gradients on the detector array.  Up to now
no suitable flat fielding procedure has been established to further
reduce the absolute photometric error.

The growth curve is used to extract spatial information.  For example,
if one calculates the growth curve of an Airy function up to its first
dark ring then the aperture radius, $r_{50\%}$, where the flux is at
50\% of its maximum value is half of the full width at half maximum
(FWHM).  Still the diffraction pattern of the telescope and the seeing
has to be subtracted.  If for an extended target one derives a 50\%
flux radius $r^e_{50\%}$ and for a PSF a radius $r^p_{50\%}$, a size
estimate of the source extension may be approximated by $
(r^o_{50\%})^2 = (r^e_{50\%})^2-(r^p_{50\%})^2$.  The size estimate by
the growth curve depends on the total flux or a normalisation at large
radii. Another diagnostic of the source extension is given by the
differential flux profile, which is a function $F(r_{i+i}) -
F(r_{i})$, and as well difficult to normalise, this time at small
radii. We take the sum of the aperture fluxes at $r \leq
r^o_{50\%}$. With such normalisation the differential flux
distribution of an extended target should be less steep than for the
PSF.

The spatial resolution of the images are at best limited by the
diffraction pattern of the 3.6m telescope, which provides a FWHM of
0.7$''$ at 10$\mu$m. In oder to increase the spatial resolution we
applied a deconvolution procedure.  We use the maximum entropy
multi--resolution wavelet decomposition by Starck \& Murtagh
(2002). The gradient algorithm has been chosen with 4 wavelet scales
assuming Gaussian distribution of the noise and we applied a $5\sigma$
threshold, where $\sigma$ is the noise as estimated on the object free
part of the image. The method provides a residual map, $R$, which is a
convolution of the deconvolved image, $D$, with the PSF, $P$, and
subtracted from the original image, $O$; symbolically: $R = D * P -
O$. The iteration of the deconvolution is stopped so that the noise
properties of the residual map is at similar level as the noise of the
original image. However, for bright point--like sources with little
extension we note that the residual map is polluted by the very bright
central peak.  Residuals are typically less than 5\% of the peak flux
of the deconvolved image.  They are caused by finite sampling of a
$\delta$--type function.  From the growth curve of the deconvolved
image we calculate its 50\% flux radius, $r^d_{50\%}$.  The residual
map is used for error propagation

To study the stability of the PSF and its influence on the
deconvolution procedure, we calculate for each PSF the deconvolution
and growth curves with any other PSF available of the same filter. For
the deconvolved PSF images we find a mean size of $\langle
r^d_{50\%}\rangle = 0.098'' \pm 0.046''$.  Size estimates from the
growth curves as measured on the raw images give a mean of $\langle
r^o_{50\%}\rangle = 0.224'' \pm 0.074''$, which already closely
matches the static aberration of the telescope. Altogether we verified
that seeing in the mid IR gives only a small contribution to the PSF
at the 3.6m. Nevertheless for deconvolution we apply the measured PSF
matched closely in time to the target observation.  A source is
regarded to be unresolved if its radius measured on the raw and
deconvolved image is below $\langle r_{50\%}\rangle +3
\sigma(r_{50\%})$. Therefore, resolved sources have $r^d_{50\%} >
0.24''$ and $r^o_{50\%} > 0.44''$, respectively.  For galactic nuclei
fulfilling these criteria, the differential source profile also shows
signatures of extended emission.

\begin{figure}
\centerline{\epsfig{figure=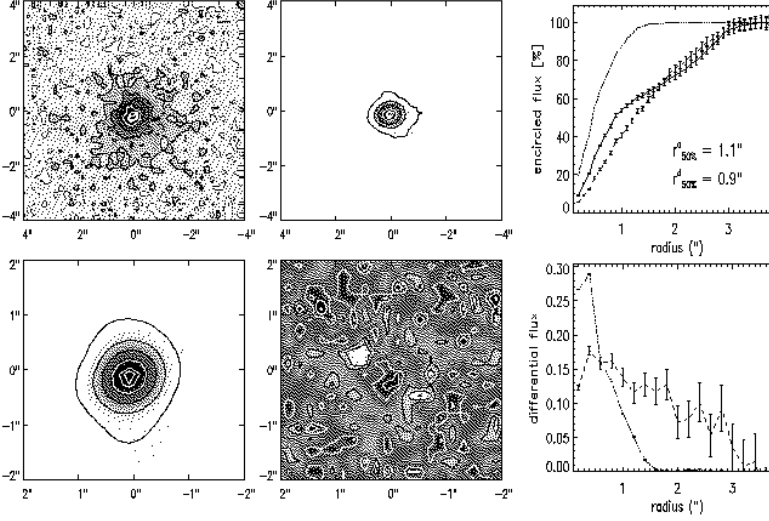,width=7.8cm,angle=90}}
\caption{NGC3256 at 10.4\,$\mu$m. On the raw (top--left), PSF
(top--middle) and deconvolved image (bottom-left) contours are in
black at 10\%, 30\% and 50\% and in white at 75\%, 90\% and 95\%
levels of the peak flux; dotted contours are at $-1\sigma$ and
$-3\sigma$ and dashed contours at $+1\sigma$ and $+3\sigma$,
respectively.  On the residual map (bottom-middle) black contours are
at $-98\%$, $-90\%$, $-50\%$, white contours at 50\%, 90\%, 98\% of
the peak. The peak fluxes and $\sigma$ values are specified in
Table~4. For all gray scale images North is up and East is
left. Growth curves (top--right) and differential flux distributions
(bottom--right) are shown for raw (dashed), PSF (dotted) and
deconvolved image (full line). Error bars are $1\sigma$ RMS.}
\label{n3256.sizefig}
\end{figure}

\begin{figure}
\caption{NGC7552 at 11.9\,$\mu$m, see Fig.~\ref{n3256.sizefig}.  On
the raw image (top--left), the PSF and the deconvolved image contours
are set in black at 8\%, 20\% and 30\% and in white at 50\%, 75\%, and
95\% levels of the peak flux}
\label{n7552.sizefig}
\end{figure}

\begin{figure}
\caption{Mrk1093 at 11.9\,$\mu$m, see Fig.~\ref{n3256.sizefig}.}
\label{mrk1093.sizefig}
\end{figure}

\begin{figure}
\caption{Circinus at 10.4\,$\mu$m, see Fig.~\ref{n3256.sizefig}.}
\label{circ.sizefig}
\end{figure}

\begin{figure}
\caption{Centaurus A at 10.4\,$\mu$m, see Fig.~\ref{n3256.sizefig}.}
\label{cena.sizefig}
\end{figure}

\begin{figure}
\caption{NGC1068 at 10.4\,$\mu$m, see Fig.~\ref{n3256.sizefig}.}
\label{n1068.sizefig}
\end{figure}

\begin{figure}
\caption{NGC1386 at 11.9\,$\mu$m, see Fig.~\ref{n3256.sizefig}.}
\label{n1386.sizefig}
\end{figure}

\begin{figure}
\caption{NGC7582 at 11.9\,$\mu$m, see Fig.~\ref{n3256.sizefig}. On
the raw image, the PSF and the deconvolved image contours are set in
black at 14\%, 20\% and 50\% and in white at 75\%, 90\%, and 95\%
levels of the peak flux, respectively.  }
\label{n7582.sizefig}
\end{figure}

\begin{figure}
\caption{Mrk509 at 11.9\,$\mu$m, see Fig.~\ref{n3256.sizefig}.}
\label{mrk509.sizefig}
\end{figure}

\begin{figure}
\caption{NGC6240 at 11.9\,$\mu$m, see Fig.~\ref{n3256.sizefig}.}
\label{n6240.sizefig}
\end{figure}

\begin{figure}
\caption{NGC7674 at 11.9\,$\mu$m, see Fig.~\ref{n3256.sizefig}.}
\label{n7674.sizefig}
\end{figure}

Observed mid--infrared flux densities for the galactic nuclei are
given in Table~3. For 10 out of 22 galaxies the nucleus is
resolved. For these we show in
Fig.~\ref{n3256.sizefig}--\ref{n7674.sizefig} raw, PSF, deconvolved
and residual maps and present growth curves as well as differential
flux distributions. Absolute levels of the contours, shown on top of
the gray scale images, can be derived together with Table~4, where we
specify the peak flux densities of the raw and deconvolved images and
the residual map. For the residual map we specify the 1$\sigma$ RMS
noise level.  The 1$\sigma$ RMS noise of the raw image is given in
col.~3 of Table~5.  It is an indicator of the sensitivity limit which
we reached with TIMMI2.

\begin{table}
\label{contour.tab}
\caption{Peak surface brightness and RMS of nuclei with extended mid
infrared emission.}
\begin{center}
\begin{tabular}{ | l | r  | r | r r | }
\hline
             &              &                   & & \\
Name         & Raw     & Deconvolved  & & Residual  \\
             & Peak     & Peak          & Peak  &  RMS \\
             &              &                   &           & \\
             &  $\frac{\rm{mJy}}{\rm{arcsec}^2}$  & $\frac{\rm{mJy}}{\rm{arcsec}^2}$  & $\frac{\rm{mJy}}{\rm{arcsec}^2}$ & $\frac{\rm{mJy}}{\rm{arcsec}^2}$ \\
             &              &                   &           & \\
\hline
             &              &                   &           & \\
Centaurus A         & 881     & 2301   & 100 & 17  \\
Circinus     & 3902    & 14806  & 167 & 31  \\
Mrk509       & 637     & 10     & 2.6 & 0.4 \\
Mrk1093      & 52      & 10     & 7.5 & 1.8 \\
NGC1068      & 10078   & 31196 & 371 & 80  \\
NGC1386      & 512     & 136   & 33  & 4.7  \\
NGC3256      & 296     & 211   & 35  & 7.3 \\
NGC6240      & 280     & 245   & 56  & 11 \\
NGC7552      & 997     & 277   & 110 & 10 \\
NGC7582      & 275     & 91    & 12  & 2.3 \\
NGC7674      & 287     & 123   & 16  & 4.0 \\
             &              &                   &           & \\
\hline
\end{tabular}
\end{center}
\end{table}

\begin{table}
\label{size.tab}
\caption{Mid--infrared sizes of galactic nuclei.}
\begin{center}
\begin{tabular}{ | l |r |c |r |r |r | }
\hline
1              & 2    & 3   & 4      & 5      & 6 \\
               &      &     &        &        &  \\
Name   & $\lambda$ & $\sigma^b$ & $r^o_{50\%}$ &  $r^d_{50\%}$ & $r^d_{50\%}$ \\
               &      &     &        &        &  \\
   & $\mu$m &$\frac{\rm{mJy}}{\rm{arcsec}^2}$ &$''$ &$''$ & pc\\
               &      &     &        &        &  \\
\hline
               &      &     &        &        &  \\
Centaurus A    & 10.4 &14   & $<$0.37 & $<$0.20& $<$3  \\
Circinus       & 10.4 &22   & 0.46   & $<$0.20& $<$4  \\
IC4329a        & 10.4 &19   &$<$0.28 & $<$0.21& $<$66  \\
IRAS05189$-$2524 &  8.6 &18   &$<$0.24 & $<$0.20& $<$170  \\
IRAS08007$-$6600 &  8.6 &66   &$<$0.23 & $<$0.27& $<$222  \\
Mrk509         & 11.9 &10   & 1.2    & 2.4    & 1642   \\
Mrk1093        & 11.9 &13   & 1.9    & 1.69   &  490  \\
NGC1068        &  8.6 &48   & 0.53   & 0.27   &   18  \\
NGC1068        & 10.4 &18   & 0.58   & 0.35   &   24  \\
NGC1365        &  8.6 &64   &$<$0.28 & $<$0.35&  $<$37  \\  
NGC1386        & 11.9 &42   & 0.50   & 0.53   &  46  \\
NGC3256        & 10.4 &28   & 1.14   & 0.88   &  157  \\
NGC4388        & 11.9 &33   &$<$0.30 & $<$0.36&  $<59$  \\
NGC5506        & 11.9 &22   &$<$0.30 & $<$0.20&  $<$24  \\
NGC6240        & 11.9 &31   &1.11    & 0.47   &  227  \\
NGC7552        & 11.9 &82   &2.79    & 2.65   &  270 \\
NGC7582        & 11.9 &17   &1.18    & 0.87   &  89  \\
NGC7674        & 11.9 &22   &0.6     & 0.46   &  264  \\
PKS2048$-$57     & 11.9 &15   &$<$0.2  & $<$0.23&  $<$51  \\
               &      &     &        &        &  \\
\hline
\end{tabular}
\end{center}

\noindent $^b 1\sigma$ RMS noise measured on the object--free part of
the raw image. \\
\noindent $^o$ Mid--infrared radius where the flux is at 50\% of the total as
derived from the growth curves of the object (raw) and PSF images. \\
\noindent $^d$ Mid--infrared radius where the flux is at 50\% of the
total as derived from the growth curve of the deconvolved image.

\end{table}

Table~5 lists the central wavelength of the TIMMI2 filter (col.~2) together
with the 50\% flux radius, $r_{50\%}$, as derived from the raw (col.~4) and
deconvolved (col.~5) image.  The latter radius is our estimate of the source
size (col.~6).

\subsection{TIMMI2 spectroscopy}

The coaddition of all chopping and nodding pairs of raw frames in
TIMMI2 spectroscopic observing mode gives two negative and one
positive long--slit spectrum on the image.  We developed an optimal
extraction procedure for TIMMI2 spectra similar to Horne (1986): A
source profile is extracted by collapsing the image along the
dispersion direction and applying a median filter.  Extension of the
sources are considered to a 2$\sigma$ cut--off, where $\sigma$ is
calculated for each column in cross--dispersion direction and on the
source free part of the array. Preliminary wavelength calibration is
performed using tables provided by the instrument team. This
calibration is fine tuned to sub-pixel accuracy by applying a linear
wavelength shift. The shift is measured as an offset to atmospheric
lines.  The procedure is applied to the target and the calibration
standard star. A division of both spectra removes the telluric lines.
To minimise residuals of the sky line cancellation, we observed
standard and target at similar airmasses.  From the standard star
spectrum we calculate the conversion factor from technical (ADU/s)
into astronomical (Jy) units. The calibrated spectra are finally
high--frequency noise filtered by the method described by Starck et
al.  (1997).  Statistical errors are computed from unfiltered
data. Because not all detector channels were always fully operational,
the spectra of M83, CenA, Circinus, IC4329A, and NGC1068 could only be
extracted long--ward of 9$\mu$m.

\subsection{ISO spectroscopy}

Whenever possible, we compare mid--infrared spectra at different
spatial scales: high--resolution long--slit TIMMI2 spectroscopy
($1''-3''$) and ISO spectra of much lower resolution ($24'' \times
24''$ with ISOPHT, Lemke et al.~1996; $14'' \times 20''$ with ISOSWS
in the N band, de Graauw et al.~1996; and ISOCAM spectra which refer,
unless otherwise stated, to the total emission of the galaxy, Cesarsky
et al.~1996).  The diffraction limits of the telescopes at 10$\mu$m
are: $\sim 0.7''$ at the ESO 3.6m and $\sim 8''$ with ISO.

For the brightest objects of our sample ISOSWS grating or line scans
are reported by Thornley et al. (2000), Sturm et al.~(2002) and Verma
et al. (2003); ISOCAM spectra by Laurent et al.~(2000), and ISOPHT
spectra by Clavel et al.~(2000), Laureijs et al.~(2000), Rigopoulou et
al.~(1999), Siebenmorgen et al.~(1999) and Spoon et al.~(2002).  For
the remaining galaxies (Cen\,A, Circinus, IC\,4329A, IRAS08007$-$6600,
Mrk509, NGC1068, NGC1365, NGC1386, NGC3256, NGC4388, NGC5506, NGC5643,
NGC6000, NGC6240, NGC7552, NGC7582) we extracted ISOPHT raw data from
the ISO archive and reduced the spectra with standard routines of the
ISOPHT interactive analysis system (PIA).  Most of the ISOPHT spectra
appear for the first time in the refereed literature.

\begin{table}
\label{bands.tab}
\caption{Important features and lines between 8 to 13$\mu$m.}
\begin{center}
\begin{tabular}{ | l | l | r | l | }
\hline
Name   &  $\lambda^a $ & $I^b$  & Mode \\
       & $\mu$m      & eV       & \\
\hline
       &             &      &       \\
PAH    &  8.6        &  $-$    & C$-$H \\
Ar~III & 8.991       & 27.6 & $-$\\
Mg~VII & 9.009       & 186.5& $-$ \\
Fe~VII & 9.537       & 99.1 & $-$\\
H$_2$~0-0~S(3) & 9.660 &  $-$ &  $-$\\
Silicate  & 9.7      &  $-$    & Si$-$O  \\
S~IV   & 10.511      & 34.8 & $-$\\
PAH    & 11.3        &  $-$    & C$-$H \\
Ne~II  & 12.814      & 21.6 & $-$\\
PAH    & 12.7        &  $-$    & C$-$H \\
       &             &   &  \\
\hline
\end{tabular}
\end{center}

\noindent $^a$ Rest wavelength.

\noindent $^b$ Ionisation energy with respect to the next lower
ionisation level.
\end{table}

\section{Results}

Here we present our high resolution mid--infrared imaging and
spectroscopy data using TIMMI2 together with the low resolution ISO
spectra.  The data of the individual objects are briefly discussed and
whenever necessary compared to observations available at other
wavelengths. We will first describe the starbursts and then the
Seyferts.

A variety of broad band dust features and fine structure lines fall
within the N band; they are summarised in Table~6.  Gas emission lines
imply either the presence of ionising shocks (Contini \& Contini 2003)
or a hard photon flux. For ionisation potential below $\sim$50eV,
photo--ionisation by hot stars is the most likely explanation for the
excitation mechanism.  In the 8--13$\mu$m spectral range there are no
lines which require very hard radiation only produced by AGNs
($>200$\,eV).  At the resolving power of TIMMI2 ($\Delta \lambda /
\lambda \sim 400$) the lines are generally unresolved. The [Ar~III]
and [Mg~VII] line occur at the same wavelength but their excitation
requirements are very different. The [Fe~VII] line may be blended by
the Ozone band, [S~IV] is in a clean atmospheric region. The PAH band
emission at 12.7$\mu$m is close to the [Ne~II] at 12.8$\mu$m.

However, as we will demonstrate by model calculations in
Sect.~\ref{models}, detection or non--detection of PAH features can
serve as a diagnostic tool to distinguish the dominating excitation
within the nucleus: starbursts or AGN; although the dividing line is
not sharp.  The mid IR spectra of spirals show dust continuum and
strong PAH bands.  Close to massive stars there is additional hot dust
continuum emission and some photo-destruction of PAH (Siebenmorgen
1993). So that starburst galaxies still show PAH bands but at lower
PAH band--to--continuum ratios than observed in spirals.  In the harsh
environment close to an AGN one expects much lower PAH
band--to--continuum ratios than in starbursts and most often PAH bands
are absent. In the immediate environment of the hard X-ray and intense
UV radiation of an AGN the PAH get destroyed. In addition significant
emission by hot dust from the inner torus wall of the unified model is
expected which further decreases the PAH band--to--continuum ratios.

The silicate band if present in the galactic nuclei is seen so far
only in absorption. From its depth one can estimate the visual
extinction. We apply $\tau_{\rm V} \sim 18 \cdot \tau_{\rm 9.7\mu
m}$. However, $\tau_{\rm 9.7\mu m}$ is difficult to estimate without
detailed model calculations of the PAH contribution to the
mid--infrared continuum (Siebenmorgen et al. 2001). Whenever possible
we interpolate a line to the spectra from $\sim$ 8 to 11 $\mu$m and so
derive a crude estimate of $\tau_{\rm 9.7\mu m}$ and the total visual
extinction. This procedure maybe uncertain within a factor 2.

\begin{figure*} 

\caption{Mid--infrared flux densities (in Jy) of  {\it starburst}
galaxies: Data presented are: ISOPHT (24$''$ aperture, histogram),
TIMMI2 spectroscopy (3$''$, full line) with 3$\sigma$ errors (shadow)
and TIMMI2 photometry with 1$\sigma$ error bars (Table~3). }
\label{sb.fig} 
\end{figure*}

\subsection{Starburst galaxies: imaging and  spectroscopy}

\subsubsection{NGC6000} 

ISO observations and radiative transfer models for the dust in
NGC~6000 are presented by Siebenmorgen et al. (1999).  At 12$\mu$m,
the flux from IRAS is stronger than from ISOPHT which suggests that
the mid IR source is extended beyond 24$''$.  TIMMI2 images at 8.6 and
10.4$\mu$m are fuzzy, without a point source.  Both spectra in
Fig.~\ref{sb.fig} are dominated by PAH emission.  In the TIMMI2
spectrum, the 11.3$\mu$m PAH band--to--continuum ratio is about two
times smaller (Table~8).  The [Ne~II] line is detected, but the
12.7$\mu$m PAH band is seen only marginally.  We estimate from the
depth of the silicate absorption in the ISOPHT (TIMMI2) spectrum a
visual extinction of $\sim 25$\,mag (20 mag); both values are
consistent with $A_{\rm V} =29$\,mag inferred from the radiative
transfer models.

\subsubsection{Mrk1466} 

The signal-to-noise in the TIMMI2 image is low, hence we cannot determine the
source size.  The only feature in the TIMMI2 spectrum detected at high
confidence is the 11.3$\mu$m PAH band (Fig.~\ref{sb.fig} and Table~8).  ISO
spectra are not available for comparison.

\subsubsection{NGC3256} 

The brightness in the TIMMI2 image increases towards the center.  The
emission is resolved and there is no point source in the core
(Fig.~\ref {n3256.sizefig}).  Both ISOPHT and TIMMI2 spectra are
dominated by PAH emission (Fig.~\ref{sb.fig}).  The 11.3$\mu$m PAH
band--to--continuum ratio is lower in the smaller aperture spectrum by
about a factor 2 (Table~8). There is strong [Ne~II] emission
(Table~7). Due to the 8.6$\mu$m PAH band and because the atmospheric
ozone band at $\sim 9.5\mu$m has not been fully eliminated, it is
difficult to say whether there is silicate absorption at all.

\subsubsection{NGC7552} 

The deconvolved 11.9$\mu$m TIMMI2 image at 0.45$''$ angular resolution
(Fig.~\ref {n7552.sizefig}) shows four hot spots, equidistant from
the center, that form a ringlike circumnuclear starburst.  Two of the
hot spots have already been detected by Schinnerer et al.~(1997) in
their 10.5$\mu$m map with 0.8$''$ resolution.  These authors also
present near infrared images and spectral synthesis models consistent
with the idea of a starburst ring.  The slit during the TIMMI2 grating
observations was positioned at the brightest knot, $\sim 3''$ North of
the centre; the [Ne~II] line is visible as well as the 11.3$\mu$m PAH
band (Table~7).  The ISOPHT spectrum shows the [Ar~II] line and strong
PAH emission.  The 11.3$\mu$m PAH band--to--continuum ratio is a
factor $\sim$3 lower in the smaller aperture spectrum (Table~8).
Silicate absorption is weak in both spectra, it corresponds to A$_{\rm
V} \sim 3$\,mag (Fig.~\ref{sb.fig}).

\begin{figure} 
\caption{Mid--infrared flux densities (in Jy) of M83: Data presented
are: ISOPHT (24$''$ aperture, histogram), TIMMI2 spectroscopy (3$''$,
full line) with 3$\sigma$ errors (shadow) and TIMMI2 photometry with
1$\sigma$ error bars (Table~3). For comparison we show the spectrum of
the Galactic Center (dashed).}\label{m83_t2.specfig}
\end{figure}

\begin{table*}
\label{lineflux.tab}
\caption{Integrated atomic line fluxes and 3$\sigma$ upper limits
($10^{-20}$W/cm$^2$) of TIMMI2  and ISOSWS  spectra.}

\begin{center}
\begin{tabular}{ | l | l | l | l | l |}
\hline
           &               &                    &               & \\
Name       &  [S IV]       & [S IV]             & [Ne II]  & [Ne II]  \\
           &               &                    &               & \\
Aperture   &  3$''$        & $14'' \times 20''$ & 3$''$    & $14'' \times 27''$\\
           &               &                    &               & \\
\hline
           &               & & &\\
Centaurus A        & $<$2.4	    & 1.4$^b$  & 22.6 $\pm$ 1.4  & 22.1$^b$ \\
Circinus	   & $<$3.6         & 12.7$^b$ & 6.6 $\pm$ 0.9    & 90$^b$ \\
IRAS05189$+$2524   & $<$1.7         &          &              & \\
IRAS08007$-$6600   & $<$1.2         &          &                   & \\
IC4329a            & 1.8 $\pm$ 0.5  &          & 1.6 $\pm$ 0.3    & \\
M83	           & $<$0.3         &          & 1.7 $\pm$ 0.2    & \\
Mrk1093	           & $<$2.2         &          & 3.0 $\pm$ 0.5    & \\
Mrk1466	           & $<$1.8         &          & $<$2.3           & \\
Mrk509 	           & $<$1.3         &          &  $<$1.3          & 1.3$^b$ \\
NGC~1068           & 30.1 $\pm$ 6.2 & 58$^b$   &   $<$14.2        & 70$^b$ \\
NGC~1365 	   & $<$1.7         & 2.6$^b$  & $<$1.8           & 40.9$^b$ \\
NGC~1386 	   & $<$2.8         &          &  $<$2.8          & \\
NGC~3256	   & $<$1.6         &  0.9$^c$ &  28.5 $\pm$ 1.0  & 89.2$^c$ \\
NGC~4388	   & 3.4 $\pm$ 0.9  &          & 5.4 $\pm$ 0.9    & \\
NGC~5506	   & 3.5 $\pm$ 0.9  &  5.4$^b$ & 2.9 $\pm$ 0.9    & 5.9$^b$ \\
NGC~5643 	   & $<$1.3         &          & 1.6 $\pm$ 0.4    &     \\
NGC~6000	   & $<$2.0         &          & 15.1 $\pm$ 1.1   & \\
NGC~6240           & $<$1.4         &          & 5.0 $\pm$ 1.6    & 17$^a$ \\
NGC~7552	   & $<$8.3         &  0.3$^c$ & 49.0 $\pm$ 5.6   & 68$^c$ \\
NGC~7582 (1.2$''$) & $<$3.7         &  1.8$^b$ & $<$8.8           & 14.8$^b$ \\
NGC~7582 (3.0$''$) & $<$1.6         &  1.8$^b$ & 8.9 $\pm$ 1.2    & 14.8$^b$ \\
PKS2048$-$57       & $<$2.7         &          & $<$3.6           & 2.1$^b$ \\
       &           &        & &          \\
\hline
\end{tabular}
\end{center}

\hspace{4cm} $^a$ Thornley et al. (2000), $^b$ Sturm et al (2002), $^c$ Verma et al. (2003)

\end{table*}

\subsubsection{Mrk1093} 

The source at 11.9$\mu$m is extended in the East--West direction
(Fig.~\ref {mrk1093.sizefig}).  TIMMI2 spectroscopy
(Fig.~\ref{sb.fig}) reveals the [Ne~II] line and PAH bands at 8.6
and 11.3 $\mu$m.  ISO spectra are not available for comparison.

\subsubsection{IRAS08007$-$6600} 

This is one of the most distant galaxy in our sample (170\,Mpc).  The
TIMMI2 images at 8.6 $\mu$m and 11.9 $\mu$m are unresolved.  The
energy distribution peaks near 60$\mu$m (Vader et al. 1993).  The
TIMMI2 spectrum does not show evidence for the presence of 8.6\,$\mu$m
PAH emission. This feature is, however, present in the larger aperture
ISOPHT spectrum (Laureijs et al. 2000). The absence may be due to low
signal-to-noise of our TIMMI2 spectrum.  The silicate absorption is
weak and indicates $A_{\rm V} \sim 6$\,mag from the central 200pc
(Table~5).

\subsubsection{M83} 

M83 is the nearest starburst galaxy in our sample. TIMMI2 images at
10.4 and 11.9\,$\mu$m show a fuzzy structure without striking point
sources.  As typical for starburst galaxies, the large aperture ISOPHT
spectrum is dominated by PAH emission bands. In contrast, the TIMMI2
spectrum of the central 60\,pc does not show any sign of PAH emission
features and is dominated instead by a smooth continuum with a weak
[Ne~II] emission line (Fig.~\ref{m83_t2.specfig}).  The shape of the
continuum is consistent with the profile of the silicate absorption
feature as found towards the ISOSWS position Sgr\,A$^*$ in the
Galactic center (Lutz 1999).  Given their similarity, we speculate
that the conditions in the nucleus of M\,83 may be similar to those in
the Galactic center. In the latter, PAHs are thought to be destroyed
by the local intense radiation field ($\sim$ 10$^{6.5}\times$ ISRF)
contributed by massive stars found in the Galactic center (Lutz 1999).
Indeed far infrared spectroscopic images of fine structures lines in
M83 indicate a strong central starburst activity headed by O9 stars
(Stacey et al. 1998).  So that this is our best example where PAH
destruction occurs by the strength of the radiation field rather than
by its hardness. The latter process is discussed in Sect.~6.  From
the depth of the silicate feature we estimate a visual extinction
$\sim$20mag.

\subsection{Seyfert galaxies: imaging and  spectroscopy}

\begin{figure*} 
\caption{Mid--infrared flux densities (in Jy) of {\it Seyfert}
galaxies: Data presented are: ISOPHT (24$''$ aperture, histogram),
ISOCAM (total field or nuclear spectrum, dashed), ISOSWS (14$\sim
20''$, dotted), UCL (5$''$, diamonds), TIMMI2 spectroscopy (3$''$,
full line) with 3$\sigma$ errors (shadow) and TIMMI2 photometry with
1$\sigma$ error bars (Table~3).}
\label{sy.fig} 
\end{figure*}

\subsubsection{Circinus}

Circinus is the closest galaxy of our sample that contains an obscured
Seyfert nucleus.  In the TIMMI2 10.4$\mu$m image
(Fig.~\ref{circ.sizefig}), half the flux comes from an unresolved
core, the other half from a 1$''$--sized blob elongated North--South.
At a comparable scale, resolved emission at 8.5$\mu$m is reported by
Siebenmorgen et al.~(1997).  The photometry of ISOPHT (this work) and
ISOSWS (Moorwood et al.~1996) give similar fluxes for $\lambda
>9\mu$m, but ISOSWS is systematically weaker 20\%--30\% between
$6-9\mu$m.  From the silicate absorption feature we estimate a visual
extinction $A_{\rm V} > 15$\,mag.

In the large aperture ISOSWS spectrum, one sees the silicate
absorption feature, strong line emission ([S~IV], [Ne~II] ) and PAH
bands.  In the TIMMI2 spectrum, [S~IV] and [Ne~II] are not detected so
they cannot be of nuclear origin.  The TIMMI2 absolute fluxes agree
well with Roche et al.~(1991) made with a 5$''$ aperture.

Moorwood \& Oliver (1994) find that the central 6pc region is
dominated by an AGN whereas star formation, as traced by H$\alpha$ and
Br$\gamma$ recombination lines, occurs in Circinus in a (partial) ring
of 200pc radius.  This ring lies outside our map of Fig.~3.  Model
calculations of the dust re-processing of the AGN are presented by
Siebenmorgen et al.~(1997) and Ruiz et al.~(2001).

Generally speaking, when one observes a Sy1 galaxy with a large
aperture, one has a free view to the inner wall of the torus and
detects strong mid--infrared emission from hot (large) grains.  The
PAH band--to--continuum ratio is then relatively low.  In a Sy2, like
Circinus, the inner torus is obscured and hence, the hot grains are
invisible and the PAH band--to--continuum ratio is higher.  This
picture is supported in the case of Circinus, but also for Sy1 and
other Sy2 galaxies (see below).

\subsubsection{Centaurus A}

This is the closest example of an AGN in our sample. In the TIMMI2
image it shows an unresolved core of less than 0.5$''$ (Fig.~\ref
{cena.sizefig}).  The raw image and the growth curve
(Fig.~\ref{cena.sizefig}) indicate weak extended emission over $\sim
2''$ with a surface brightness ten times lower than at the peak.  The
ISOCAM (Laurent et al.~2000) and ISOPHT (this work) spectra are PAH
dominated whereas the TIMMI2 spectrum is featureless except for the
presence of a strong [Ne~II] line (Fig.~\ref{sy.fig}). The visual
extinction estimated from the silicate absorption detected in the
ISOPHT and TIMMI2 spectra is around 14\,mag.

\subsubsection{NGC1068} 

TIMMI2 resolves the nucleus of this prototype Sy2 galaxy (Fig.~\ref
{n1068.sizefig}) and our image of the core agrees with the long
integrations by Alloin et al.~(2000) using CAMIRAS and Bock et
al. (2000) with MERLIN at Keck.  However, we are not sensitive enough
to confirm the individual hot cores found in those maps. PAHs are
detected in the integrated ISOCAM (Laurent et al.~2000) and in our
ISOPHT spectrum, but in a smaller aperture they disappear (Fig.~\ref
{sy.fig}, see also the ISOSWS spectrum of Lutz et al.~2000).  The
continuum is of similar strength for ISOSWS and TIMMI2 indicating that
the AGN is indeed the major source in the central 1kpc region.

The ISOCAM spectrum of the central $9''$ by Laurent et al.~(2000) is,
like ISOSWS and TIMMI2, free of PAHs.  The [Ne~II] line is absent in
TIMMI2, but strong in the ISOSWS spectrum, [S~IV] is seen in both.
The strength of the silicate absorption band increases with spatial
resolution; for the central 20pc, we estimate $A_{\rm V} \ge 5$\,mag.

\subsubsection{NGC5643} 
 
This Sy2 galaxy shows high-excitation optical lines (Kinney et
al.~1993).  The emission in [O~III] and H${\alpha}$ is extended up to
$\sim 20''$ (Schmitt et al. 1994).  The 11.9$\mu$m TIMMI2 image is
fuzzy, without a point source. The [Ne~II] line is detected by TIMMI2
and for the [S~IV] line we get an upper limit (Table~7).  PAHs are
strong in the ISOPHT aperture, whether they are seen with TIMMI2 is
unclear (Fig.~\ref{sy.fig}).

\subsubsection{NGC1386} 

The TIMMI2 image of this Sy2 galaxy (Ferruit et al.~2000) has an
unresolved core (Table~5) and an extended component
(Fig.~\ref{n1386.sizefig}).  The ISOPHT spectrum shows PAH bands and
tentative, weak silicate absorption. In the TIMMI2 spectrum,
(Fig.~\ref{sy.fig}) one sees the silicate feature, corresponding to
$A_{\rm V} \approx 30$\,mag, but no PAHs.

\subsubsection{NGC1365} 

A face--on Sy1 galaxy with an unresolved core in the TIMMI2 image.
Its mid--infrared emission is dominated by PAHs everywhere except in
the very center.  The band--to--continuum ratio decreases with
decreasing aperture size (Fig.~\ref {sy.fig}). PAH bands are not
present in the TIMMI2 spectrum (Table~8).  [S~IV] and [Ne~II] emission
is detected only with ISOSWS, but not with TIMMI2.  The TIMMI2
continuum is flat without evidence of silicate absorption.


\subsubsection{NGC4388} 

A galaxy with a Sy2 nucleus where Ho \& Ulvestad (2001) see at radio
wavelengths two components, ~1.9$''$ (150 pc) apart, none of them
coincides with the position of the optical peak. A hard X-ray peak is
found at the position of the active nucleus (Iwasawa et al. 2003).  As
in the previous examples, the TIMMI2 image has a single unresolved
core (Table~5), and there are strong PAH bands in the ISOPHT, but none
in the TIMMI2 spectrum (Fig.~\ref {sy.fig}).  The [S~IV] and [Ne~II]
lines are detected with TIMMI2 (Table~7).

\subsubsection{IRAS05189$-$2524} 

In this most distant object of our sample, the central source in the
TIMMI2 image is unresolved (Table~5).  The same conclusion was reached
by Soifer et al.~(2000) using MERLIN at Keck.  ISOPHT sees weak PAH
emission (Laureijs et al.~2000) and an absorption feature at 6$\mu$m
attributed to water ice (Spoon et al. 2002).  PAH bands
(Fig.~\ref{sy.fig}) as well as gas emission lines are absent in the
TIMMI2 spectrum.  From the silicate absorption, we estimate $A_{\rm V}
\sim 12$\,mag.

\begin{figure*} 
\caption{Mid--infrared flux densities (in Jy) of {\it Seyfert}
galaxies: Data presented are: ISOPHT (24$''$ aperture, histogram),
 TIMMI2 spectroscopy (3$''$, full line and 1.2$''$ dotted) with
3$\sigma$ errors (shadow) and TIMMI2 photometry with 1$\sigma$ error
bars (Table~3). }  
\label{syb.fig} 
\end{figure*}

\subsubsection{NGC7582} 

The spectrum is a superposition of a Sy2 nucleus and a starburst
(Storchi--Bergmann et al.~2000).  The AGN is responsible for X--ray
emission over a few kpc (Levenson et al.~2000). Optical and near IR
images (Regan \& Mulchaey 1999) indicate a complex dust morphology.
There is a lane that runs from the southeast to an absorbing ring
north of the chain of blue clumps; the TIMMI2 image reflects this
structure (Fig.~\ref{n7582.sizefig}).  PAH bands are detected with
ISOPHT, but not [S~IV].  The excellent seeing conditions during the
TIMMI2 observations allowed us to use besides the 3$''$ also the
1.2$''$ slit.  The spectra look quite similar at both resolutions
(Fig.~\ref{syb.fig}): [Ne~II] is present, the PAH band is detected
at 11.3$\mu$m but not at 8.6 $\mu$m.  From the silicate absorption we
estimate A$_{\rm V} \sim 20$\,mag.  The 11.3$\mu$m PAH
band--to--continuum ratio is in TIMMI2 lower than in ISOPHT by about a
factor 6 (Table~8). 

\subsubsection{NGC5506} 

The optical spectrum of this Sy2 galaxy is comparable to what one
finds in young starburst galaxies, however, the high excitation lines
([Ne~V], [Fe~VII], [He~II]) mandate the presence of a non--thermal
continuum (Oliva et al.~1999). The X--ray spectrum is highly absorbed
(Pfefferkorn et al. 2001).  In the 8--13$\mu$m wavelength range,
[Ar~III]/[Mg~VII], [S~IV] and [Ne~II] emission lines are detected with
ISOSWS, [S~IV] and [Ne~II] lines at similar strengths with TIMMI2
(Table~7).  Evidence of PAH emission is found in the ISOPHT spectrum
at 6.2$\mu$m (Rigopoulou et al.~1999), whereas TIMMI2 sees a
featureless spectrum.  TIMMI2 and ISOPHT agree very well
photometrically (Fig.~\ref{syb.fig}).  We estimate an optical depth
of over 15\,mag towards the core.

\subsubsection{PKS2048$-$57}

In the near--infrared, the nuclear source of this object is very red
(K--L = 2.1\,mag) and heavily polarised (17\%) which is explained by a
non-thermal contribution (Hough et al.~1987).  Its spectral energy
distributions peaks at 60$\mu$m (Hessler \& Vader 1995) and the radio
power is 2 orders of magnitude greater than that of typical Seyferts.
The mid IR image is dominated by an unresolved core.  The ISOPHT
spectrum shows weak PAH emission. The TIMMI2 spectrum is badly
affected by telluric sampling noise and no PAH features are visible.
[Ne~II] and [Ar~III]/[Mg~VII] is detected by ISOSWS but not with
TIMMI2.

\subsubsection{IC4329A}

This Sy1 galaxy is point--like in our TIMMI2 images, but shows
extended [O~III] and H$\alpha$ emission (Mulchaey et al.~1996).  The
TIMMI2 and ISOPHT spectra appear to trace the same components, except
for the [S~IV] line which is stronger in the large aperture
(Fig.~\ref{syb.fig}).  [Ne~II] is seen with TIMMI2; the feature near
9.4$\mu$m is probably telluric.

\subsubsection{Mrk509}

The emission of this Sy1 galaxy is at 11.9$\mu$m dominated by an
unresolved core and marginal resolved emission below the 10\% level of
the peak flux (Table~4) is found toward the South--East
(Fig.~\ref{mrk509.sizefig}). This may be supported by the slight
increase of the mid--infrared emission in the larger ISOPHT aperture
as compared to the TIMMI2 spectrum (Fig.~\ref{syb.fig}). Source
extension towards the south is also reported by near infrared
observations (Winge et al. 2000). The ISOPHT spectrum is typical for a
Sy1 (Clavel et al. 2000), it shows weak PAH emission over a strong
continuum.  In the TIMMI2 spectrum PAH bands are not detected
(Table~8). Both spectra do not show evidence of silicate absorption.
[Ne~II] is detected by ISOSWS but not with TIMMI2.

\subsubsection{NGC6240} 

A famous ultraluminous merger with a deeply hidden AGN.  The AGN
reveals itself by the strong iron line at 6.4\,keV and the powerful
continuum at 100 keV (Matt et al.~2000). Two hard X-ray nuclei have
been recently detected by Komossa et al. (2003). Near infrared spectra
indicate very low metalicity ($\sim $1/10 solar, Oliva \& Origlia
1998).  NICMOS images (Scoville et al.~2000) show two nuclei separated
by 1.6$''$ (0.8 kpc) North--South.  There are also two nuclei at radio
wavelengths 1.4$''$ apart (Condon et al.~1996), the CO (2-1) peak lies
in between (Tacconi et al.~1999). The TIMMI2 image at 11.9\,$\mu$m
(Fig. 8) reveals the nuclear region to be dominated by a strong core,
which we identify with the Southern nucleus (Tecza et al. 2000). A
fainter structure, at 10\% of the peak flux of the core, extends
$\sim$2$\arcsec$ to the North and matches the structure observed at
lower resolution at near-infrared wavelengths (Tecza et al. 2000).

The ISOPHT spectrum is dominated by strong PAH emission and shows in
addition emission lines of [Ar~II], H$_2$~0-0~S(5) (blended with
[Ar~II]) and H$_2$~0-0~S(3). Interestingly, the latter line is absent
in our TIMMI2 spectrum. Most of the [Ne~II] emission ($\sim$2/3)
originates from outside the 3$\arcsec$ slit (see Table 7). Both
spectra (Fig.~\ref{syb.fig}) show prominent PAH emission. The
$8.6\mu$m/$7.7\mu$m band ratio is small in ISOPHT and the PAH band at
8.6$\mu$m feature is not even detected with TIMMI2.  Between 9 and
10.5$\mu$m the TIMMI2 spectrum is noise dominated. The silicate
absorption, as of ISOPHT suggests A$_{\rm V} > 25$\,mag.

\subsubsection{NGC7674} 

This Sy2 is slightly extended in the radio and has within 0.5$''$ a
double source (Kukula et al. 1995).  The deconvolved 11.9$\mu$m TIMMI2
image is also extended and at 0.5$''$ (Fig.~\ref{n7674.sizefig},
Table~5) dominated by a single core.  Previous mid IR maps by Miles et
al.~(1996) have insufficient resolution to resolve the galaxy but
their absolute photometry is consistent with ours.  The 5$''$
resolution UCL spectrum (Roche et al.~1991) is rather featureless,
whereas the ISOPHT spectrum contains several PAH bands (Schulz,
priv.~communication). We could not yet perform TIMMI2 spectroscopy.

\begin{table}
\label{pahratio.tab}
\caption{Continuum subtracted $11.3\mu$m PAH band fluxes, $F$, (in
$10^{-20}$W/cm$^2$) and PAH band--to--continuum ratios, $R$, of the
TIMMI2 (3$''$ slit) and ISOPHT (24$''$ aperture) spectra.}

\begin{center}
\begin{tabular}{ | l | l | l | l | l |}
\hline
       &  & & & \\
Name & $F^{\rm{TIMMI2}}$ & $R^{\rm{TIMMI2}}$ & $F^{\rm{ISO}}$ &
$R^{\rm{ISO}}$ \\
\hline
       &  & & & \\
Circinus              &$<$50  &$<$ 0.1 &  530  & 0.4\\
Centaurus~A             &$<$11  &$<$ 0.1 &  100  & 0.9\\
IC4329a               &$<$ 6  &$<$ 0.1 &$<$10  & $<$0.1\\
IRAS05189$+$2524      &$<$10  &$<$ 0.1 &$<$ 3  & $<$0.1 \\
IRAS08007$-$6600      &$<$6   &$<$ 0.1 &$<$10  & $<$0.1\\
M83                   &$<$ 2  &$<$ 0.1 & 252   & 1.2 \\
Mrk509                &$<$ 7  &$<$ 0.1 &   17  & 0.1\\
Mrk1093               & 10    &    1.0 &$-$    & $-$\\
Mrk1466               &    4  &    1.1 & $-$   & $-$\\
NGC~1068              &$<$95  &$<$ 0.1 &$<$750 & $<$0.1\\
NGC~1365              &$<$11  &    0.2 &  125  & 0.6\\
NGC~1386              &$<$16  &$<$ 0.1 &  165  & 0.6\\
NGC~3256              & 30    &    0.4 &  107  & 1.0\\
NGC~4388              &$<$11  &$<$ 0.2 &  17   & 0.6\\
NGC~5506              &$<$16  &$<$ 0.1 &$<$15  & $<$0.1\\
NGC~5643              &$<$10  &    0.2 &  20   & 0.5\\
NGC~6000              & 16    &    0.6 &  67   & 1.1\\
NGC~6240              &   12  &    0.6 &   21  & 1.0\\
NGC~7552              & 33    &    0.3 &  162  & 0.9\\
NGC~7582              & 19    &    0.2 &  110  & 1.3\\
PKS2048$-$57          &$<$22  &$<$ 0.1 &$<$50  & $<$0.1\\
                      &       &        &       &    \\
\hline
\end{tabular}
\end{center}
\end{table}

\section{Heating and evaporation of PAHs}

We discuss the survival of PAHs in starburst and AGN environments with
energetic photons of, at least, a few hundred eV.  Such photons can
only be emitted by an X--ray source.

\subsection{The evaporation condition}

Let $P(T)$ be the temperature distribution of an ensemble of very
small grains, for example, PAHs, so that $P(T)\,dT$ gives the
probability of finding a particular grain in the temperature interval
$dT$ around $T$.  Whereas grain cooling usually proceeds through
emission of infrared photons, in a very hot particle, above
$\sim$2500\,K, evaporation is more effective (see Fig.~8 of Voit,
1991).  The evaporation rate is proportional to the number of surface
atoms, $N_{\rm surf}$, for which approximately $N_{\rm surf} \sim
N^{2/3}$ if $N$ is the total number of atoms in the grain.

\medskip
Let $P_1$ and $P_2$ be the probabilities that a surface atom is at an
energy level below or above the binding energy $E_{\rm b}$,
respectively.  Usually, $P_2 \ll 1$, so $P_1\simeq 1$, and obviously
$P_1 +P_2=1$.  For a Boltzmann distribution, $P_2/P_1 \sim
\exp(-E_{\rm b}/kT)$.  Due to phonon collisions, an atom changes its
level at a rate given by the Debye frequency $\nu_{\rm D}$.  In one
second, the atom is $P_2\nu_{\rm D}$ times above $E_{\rm b}$, the
evaporation rate is therefore
\begin{equation}\label{VerdampfRate}
R_{\rm ev} \ = \ N_{\rm surf} \, \nu_{\rm D} \, e^{-E_{\rm b}/kT} \ .
\end{equation}

Evaporation rises exponentially with temperature.  To counterbalance
it, atoms must be built into the grain.  The rate at which this occurs
is proportional to the thermal velocity of atoms, or to $T^{1/2}$, and
to the density of the interstellar medium.  The exponential process of
evaporation wins over accretion by orders of magnitude in a fairly
narrow range of $T$, the density is not relevant.  Let us assume that
the grains evaporate if a certain fraction of them, $f_{\rm ev}$, is
above some critical temperature $T_{\rm cr}$.  The latter should be
somewhat greater than $T_{\rm ev}$.  Relegating all complicated or
unknown physics to the parameters $f_{\rm ev}$ and $T_{\rm cr}$, we
write the condition for evaporation as
\begin{equation}\label{VerdampfBed}
I_{\rm ev} \ \equiv \ \int_{T_{\rm cr}}^\infty P(T)\,dT \ > \ f_{\rm ev}  \ .
\end{equation}

Reasonable numbers for $T_{\rm cr}$ in (\ref{VerdampfBed}) are 2500\,K and
$f_{\rm ev}\sim 10^{-8}$, but such suggestions do, of course, not constitute a
theory.

\begin{figure}
\caption{The absorption efficiency of a graphite sphere of 10\AA\
radius.  Mie theory is still assumed to be valid at X--rays.  Optical
constants from Laor \& Draine (1993).}
\label{QgraphXray.fig}
\end{figure}

\subsection{X--ray cross sections for dust absorption}

An AGN differs from a starburst nucleus of the same bolometric luminosity by the
hardness of its radiation field.  We are here interested in the question how the
presence of very energetic photons in AGNs affects the chance of survival of
small grains.  First, we remark briefly on the X--ray absorption cross section
of dust.

When a grain absorbs an optical or UV photon of frequency $\nu$, its
internal energy rises by the full amount $h\nu$.  Furthermore, when
the grain is not smaller than the wavelength, the absorption
efficiency is approximately one.  The situation is different at
X--rays.  The interaction of hard photons with interstellar dust and
the limits of Mie theory have been studied by Voit (1991), Laor \&
Draine (1993) and Dwek \& Smith (1996).  Below 10\,keV, the main
absorption process is photo--ionization of inner atomic shells.  The
excited electron loses energy in inelastic scattering as it travels
through the grain, but may also leave the grain and carry away kinetic
energy as a photo--electron.  When an X--ray photon has created a gap
in the innermost shell, the gap will be filled by a downward
transition of an electron from an upper shell and the energy may
escape either as a photon, or non--radiatively through ejection of an
Auger electron.

The computations of Dwek \& Smith (1996) indicate that for graphite
particles of 50\AA\ radius, soft X--ray photons with $h\nu \simless
100$\,eV deposit practically all their energy in the grain and Mie
theory is still valid.  For photon energies above $h\nu =100$\,eV and
particle radii $a < 50$\AA, only part of the photon energy is
deposited in the grain and we therefore reduce the absorption
efficiency by a factor $\propto 1/{\nu}$.

Fig.~\ref{QgraphXray.fig} illustrates the absorption efficiency,
$Q^{\rm abs}$, from Mie theory of graphite grains with radii from 10
to 1000\AA\ (see Fig.~2 of Laor \& Draine (1993) for a wider range
of parameters).  The jump at 43\AA\ (290\,eV) is due to the ionization
threshold of the K--shell of the carbon atoms.  At still higher
frequencies, one has the typical $\nu^{-3}$ cross section for ionizing
solitary atoms.  The drop in $Q^{\rm abs}$ at wavelengths short of the
first ionization level at $\sim$10\,eV is less steep ($Q^{\rm abs}
\propto \nu^{-2}$ for $a\le100$\AA); for big grains ($a>1000$\AA),
$Q^{\rm abs}$ is essentially flat at these wavelengths.

When, as for X--rays, absorption is due to the excitation of inner
atomic shells, the cross section of a grain, $K^{\rm abs}$, depends
only weakly on the way the atoms are held together in the solid.  One
can therefore roughly set $K^{\rm abs}$ proportional to the number of
atoms in the grain.  The dust particle in Fig.~\ref{QgraphXray.fig}
with 10\AA\ radius consists of $N_{\rm C}=520$ carbon atoms and has
$Q^{\rm abs} \simeq 0.2$ in the interval 10eV $\le h\nu\le 20$eV, and
$Q^{\rm abs} \propto \nu^{-2.2}$ for $h\nu > 20$eV.  For the cross
section per carbon atom, $C^{\rm abs}$, this implies $C^{\rm abs}
\simeq 10^{-17}$\,cm$^2$ for 10eV $\le h\nu\le 20$eV, and $C^{\rm abs}
\simeq 10^{-17}\times (20/h\nu)^{2.2}$\,cm$^2$ for $h\nu > 20$eV.  We
will use these crude numbers to calculate the cross section of PAHs.

\begin{figure}
\caption{The distribution function of the temperature, $P(T)$, for
graphite grains of 5\AA\ radius.  Monochromatic flux $F=10^4$\,erg
cm$^{-2}$ s$^{-1}$, photon energies from 10 to 80eV. }
\label{PvonT}
\end{figure}

\subsection{The temperature distribution of PAHs in a hard radiation field}

To investigate how the distribution function $P(T)$ and, in
particular, the evaporation condition (\ref{VerdampfBed}) depend on
the photon energy, we consider a source with a monochromatic flux of
strength $F$, all photons have then the same energy $h\nu$.  While
keeping the value of the flux fixed, we change $h\nu$, so if $N_{\rm
ph}$ is the number of photons, $N_{\rm ph} h\nu = F$ stays constant.
Low energy photons impinge more frequently, but raise the enthalpy $H$
of the grain only by a small amount.  X--ray absorption, on the other
hand, is a relatively rare event, but accompanied by a big jump in
$H$.  Temperature distributions $P(T)$ are shown in Fig.~\ref{PvonT}
for a flux $F=10^4$ erg cm$^{-2}$ s$^{-1}$ corresponding to a
luminosity $L=2.4\times 10^{11}$ L$_\odot$ at a distance of 100\,pc.
The photon energies vary between 10 and 80eV.  Grains much above the
evaporation temperature of the solid ($\sim$2000\,K) are, of course,
unrealistic.

\begin{figure}
\caption{The fraction of graphite grains of 5\AA\ (solid lines) and
10\AA\ (dashed lines) radius above a critical temperature $T_{\rm cr}
= 2500$\,K (see Equation~\ref{VerdampfBed}).  The strength of the
monochromatic flux $F$ is indicated. }
\label{vsg_evap}
\end{figure}

\begin{figure}
\caption{The fraction of PAHs with $N_{\rm C} = 100$ carbon atoms
above a critical temperature $T_{\rm cr} = 2500$\,K (see Equation~\ref
{VerdampfBed}).  The strength of the monochromatic flux $F$ is
indicated.  }
\label{pah_evap}
\end{figure}

The distributions $P(T)$ in Fig.~\ref{PvonT} display a kink at the
temperature where the thermal energy of the grain equals $h\nu$.
Still higher grain (momentary) temperatures are achieved when a second
photon is absorbed before the grain has cooled off from the excitation
by the first.

Fig.~\ref{vsg_evap} displays the integral $I_{\rm ev}$ in equation
(\ref {VerdampfBed}) as a function of photon energy $h\nu$ for various
fluxes and two grain sizes.  The important point is here that for the
smallest graphite grains with 5\AA\ radius (65 carbon atoms), $I_{\rm
ev}$ grows very rapidly in the range from 20 to 40eV and then declines
only gradually.  The energy of photons, $\langle h\nu\rangle$, emitted
by the hottest O stars ($T_{\rm eff}=5\times 10^4$\,K) is only 16\,eV
when averaged over the total spectrum, and equals 22\,eV when averaged
over the Lyman continuum ($\lambda < 912$\AA).  These numbers are thus
upper limits for the radiation field in a starburst.  Furthermore, the
stellar flux of an O star plummets at wavelengths below 228\AA, or
above $h\nu = 54.1$\,eV, when helium becomes doubly ionized.  AGNs, on
the other hand, have a spectrum typically declining with $\nu^{-0.7}$
and emit copious amounts of extreme UV and soft X--ray photons as
witnessed by the observation of fine structure lines from ions with
high ($> 100$\,eV) ionization potentials.

We conclude from Fig.~\ref{vsg_evap} that small graphite grains
(5\AA\ radius) survive in a starburst as long as fluxes are below
$\sim 10^4$\,erg cm$^{-2}$, but they will be destroyed near an AGN of
the same flux $F$.  The fate of PAHs of equivalent mass is similar
(see Fig.~\ref{pah_evap}).  Bigger grains (see curves for 10\AA\
radius in Fig.~\ref{vsg_evap}) are much more resistive.  They need
higher fluxes to get vaporized and survive in a starburst environment.

\section{Radiative transfer models}
\label{models}

Here we present comprehensive model calculations for the mid--infrared
emission of starburst and AGN radiation environments in the optical
thin and optical thick regime.

\begin{figure}
\caption{Optically thin dust emission (full line) between 5 and
15\,$\mu$m in a starburst (top) and an AGN (bottom) at various
distances to the central heating source. The total luminosity is $L =
10^{11}$\Lsun .  The emission of large grains, representing the
continuum, is shown dashed. }
\label{sbagn_thin.fig}
\end{figure}

\subsection{Optically thin dust emission}

In a first attempt to understand and to illustrate the influence of
the hardness of the radiation field, we compute the optically thin
dust emission at various distances from the source.  A point--like
heating source of total luminosity $L = 10^{11}$\Lsun\ has either the
spectrum of an OB star ($T_* = 25\,000$\,K) to mimic a starburst, or
of an AGN with $L_\nu \propto \nu^{-0.7}$ in the range from 10\AA \/
to $\sim$2$\mu$m (the exact upper limit is unimportant as the dust has
little extinction there).  The distance to the source, and thus the
flux which the dust receives, is varied from 4 to 100\,pc.

For simplicity we consider in this section only two dust components:
Large carbon grains with a $n(a) \propto a^{-3.5}$ size distribution
and radii between 300 and 2400\AA, and PAHs consisting of 100 C and 28
H atoms and an abundance of 10\% of the large carbon grains.  The
spectra are shown in Fig.~\ref{sbagn_thin.fig}.  They are in arbitrary
units and normalised at the maximum.

One notices that large grains become hotter as the radiation field
hardens.  At a distance of 4\,pc to the heating source, PAHs
invariably evaporate, at 10\,pc evaporation occurs only near an AGN,
and at 100\,pc they survive in both environments, starburst and AGN.

\subsection{Optically thick dust emission}

\begin{figure}
\centerline{\hspace{1cm}\epsfig{figure=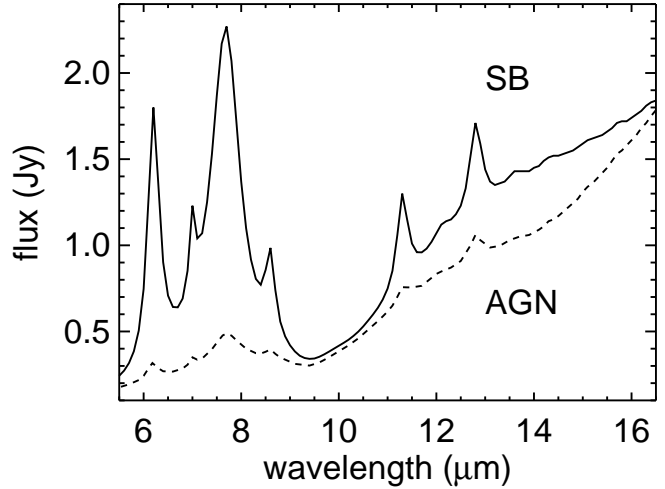,width=7.5cm,angle=90.}}
\caption{Comparison of radiative transfer model spectra of dusty
starbursts (solid) and AGN (dashed).  In both models the visual
extinction towards the cloud centre is $A_{\rm V} = 25$\,mag and
the luminosity is $L = 10^{11}$\Lsun. Spectra are calculated for a
distance of 50\,Mpc. }
\label{sb_agn_av25.fig}
\end{figure}

\begin{figure}
\centerline{\hspace{1cm}\epsfig{figure=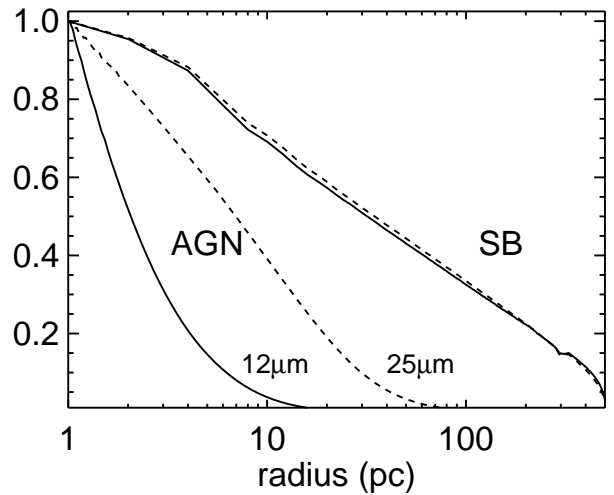,width=7.5cm,angle=90}}
\caption{Normalised intensities at 12 (solid lines) and 25$\mu$m
(dashed lines) for starbursts and AGN models as presented in
Fig.~\ref{sb_agn_av25.fig}. At both wavelengths the predicted FWHM
of the starburst galaxy is much larger than for the AGN.}
\label{av25_cuts.fig}
\end{figure}

Nuclei of luminous infrared galaxies are generally dust enshrouded and
not transparent.  Radiative transfer calculations of optically thick
dusty nuclei have been carried out in various approximations (Kr\"ugel
\& Tutokov 1978, Rowan-Robinson \& Crawford 1989, Pier \& Krolik 1993,
Loar \& Draine 1993, Granato \& Danese 1994, Kr\"ugel \& Siebenmorgen
1984, Siebenmorgen et al. 1997, Silva et al. 1998, Efstathiou et
al. 2000, Ruiz et al.~2001, Siebenmorgen et al. 2001, Nenkova et
al. 2002).

In the subsequent spherical models, for starbursts and AGNs, the total
luminosity is again $10^{11}$\Lsun.  The dust cloud has a constant
density, a visual extinction of $A_{\rm V} \sim 25$\,mag (from the
surface to the centre), and an outer cloud radius of 500\,pc.  To be a
bit more realistic the dust consists here, besides large carbon
grains, also of large silicate grains of the same size distribution
and with a 12\% larger total volume.  Absorption and scattering
efficiencies are calculated using Mie theory and optical constants by
Loar \& Draine (1993).  In addition, there are small graphites and
small silicates of 10\AA\ radius, with an abundance of 0.5\% that of
the large carbon grains. PAH parameters are as described in the
previous section.

The radiative transfer in a dust cloud around a starburst is computed
following Siebenmorgen et al.~(2001), around an AGN after Siebenmorgen
et al. (1997) and the inclusion of soft X-rays as in this work.  The
major difference between both radiative transfer models is that AGN
are heated by a central source whereas starburst galaxies are heated
by stars which are distributed over a large volume.  In case of a
starburst, each of the OB stars have a uniform luminosity of $20\,000$
\Lsun\ and a stellar temperature $T_* = 25\,000$\,K. The stellar
density changes with galactic radius $r$ up to 500\,pc from the center
like $\propto 1/r$ (there must be a cutoff at $r=0$, but its exact
position is unimportant).  Each OB star is surrounded by a {\it hot
spot} of constant dust density, corresponding to $n({\rm H}) \sim
10^4$\,cm$^{-3}$.  The size of the hot spot is determined by the
condition of equal heating of the dust from the star and from the
radiation field in the galactic nucleus.  As for the AGN models, the
inner radius of the hot spots is given by the photo--destruction or
evaporation radius of the grains.

Our model spectrum for the starburst and the AGN is displayed in
Fig.~\ref {sb_agn_av25.fig}.  The starburst shows strong PAH
emission and a continuum which rises steeply beyond 10$\mu$m, whereas
the AGN is almost free of PAH emission.  Cross cuts over the model
nuclei, at various wavelengths, are shown in
Fig.~\ref{av25_cuts.fig}.  At 12$\mu$m and 25$\mu$m, the AGN is much
more compact than the starburst.  Therefore the mid--infrared size may
be used to distinguish between the two kinds of energy sources. This
statement holds unless the stars are distributed by more than a few
pc.

\section{Conclusion}

We presented mid infrared imaging and spectroscopy for a sample of 23
galactic nuclei. The observations were obtained with the ISO ($14'' -
24''$ resolution) and with TIMMI2 at the 3.6m ESO telescope ($3''$ and
$0.5''$ resolution for spectroscopy and imaging, respectively).  The
galaxies are mostly starbursts and AGNs, with apparent mid--infrared
fluxes of 20 mJy or more.  Our main results are:

{\it a)} Seyfert nuclei are dominated by a compact core while
starbursts do not display such a central concentration; they are
usually spatially resolved by TIMMI2.

{\it b)} Almost all our ISO spectra (18 out of 19) are marked by bands
at 6.2, 7.7, 8.6 and 11.3$\mu$m, generally attributed to PAHs.  In the
TIMMI2 spectra, which refer to the innermost nuclear regions, these
bands are present in starbursts, but not detected or significantly
reduced in AGNs.  This corroborates previous suggestions that PAHs
evaporate much easier in the hard photon environment of an AGN, but
are often able to survive in a starburst.

{\it c)} Applying a simple criterion for dust evaporation, we quantify
the conditions under which small grains are photo--destructed in a
monoenergetic hard radiation field. Analyzing their temperature
distributions, we find that PAHs or very small graphite particles with
less than 100 carbon atoms cannot survive if the photons have energies
above 20\,eV and their flux is greater than $10^3$\,erg cm$^{-2}$
s$^{-1}$.  The detection or non--detection of PAH bands can thus serve
as a diagnosis whether a galaxy is predominantly powered by an AGN or
a starburst.  Of course a very dense stellar cluster provides a high
radiative density and destroys PAH (Siebenmorgen 1993). Therefore the
dividing line between AGN and starbursts is not sharp, as demonstrated
by the nuclear spectrum of M83 and Circinus.

{\it d)} In radiative transfer models that include soft X--rays and
incorporate the above dust evaporation criterion, the starburst
galaxies show PAH bands everywhere, and in the mid--infrared they are
stronger and more extended than AGNs of the same luminosity.  

{\it e)} The [S~IV] and [Ne~II] fine structure lines at 10.5 and
12.8$\mu$m, which can still be excited in an O star environment, are
in both types of galaxies generally stronger in the large (ISO)
aperture spectra indicating that their emission comes from outside the
nucleus.

\acknowledgements 

PIA is a joint development by the ESA Astrophysics Division and the
ISOPHT Consortium. This research has made use of the NASA/IPAC
Extragalactic Database (NED) which is operated by the Jet Propulsion
Laboratory, California Institute of Technology, under contract with
the National Aeronautics and Space Administration.

\end{document}